\documentclass[twocolumn]{aastex62}
\usepackage{CJK}
\usepackage{amsbsy}
\usepackage{multirow}
\usepackage{bm}

\def\msun{\,{\rm M}_\odot}

\def\mpch{\,{\rm h^{-1}  Mpc}}

\def\bx{{\boldsymbol x}}

\newcommand{\pks}{P_{\rm KS}}
\newcommand{\pku}{P_{\rm KU}}
\newcommand{\nsate}{\rm N_{sate}}
\newcommand{\meancosangle}{\langle\cos(\theta)\rangle}
\newcommand{\meanangle}{\langle\theta\rangle}

\received{XXX}
\revised{YYY}
\accepted{ZZZ}

\shorttitle{satellite ellipsoid  in filament}
\shortauthors{Peng Wang et al.}

\begin{document}
\begin{CJK*}{UTF8}{gbsn}

\title{The alignment of satellite systems with cosmic filaments in the SDSS DR12}
\correspondingauthor{Peng Wang}
\email{pwang@aip.de}

\author{Peng Wang (王鹏)}
\affil{Leibniz-Institut f\"ur Astrophysik Potsdam, An der Sternwarte 16, D-14482 Potsdam, Germany}

\author{Noam I. Libeskind}
\affil{Leibniz-Institut f\"ur Astrophysik Potsdam, An der Sternwarte 16, D-14482 Potsdam, Germany}
\affil{University of Lyon; UCB Lyon 1/CNRS/IN2P3; IPN Lyon (IPNL), France}

\author{Elmo Tempel}
\affil{Tartu Observatory, University of Tartu, Observatooriumi 1, 61602 T\~oravere, Estonia}

\author{Marcel S. Pawlowski}
\affil{Leibniz-Institut f\"ur Astrophysik Potsdam, An der Sternwarte 16, D-14482 Potsdam, Germany}

\author{Xi Kang (康熙)}
\affil{Zhejiang University-Purple Mountain Observatory Joint Research Center for Astronomy, Zhejiang University, Hangzhou 310027, China}
\affil{Purple Mountain Observatory, No. 10 Yuan Hua Road, 210034 Nanjing, China}

\author{Quan Guo (郭铨)}
\affil{Shanghai Astronomical Observatory, Nandan Road 80, Shanghai 200030, China}

\begin{abstract}
Galaxies, as well as their satellites, are known to form within the cosmic web: the large, multi-scale distribution of matter in the universe. It is known that the surrounding large scale structure (LSS) can impact and influence the formation of galaxies, e.g. the spin and shape of haloes or galaxies are correlated with the LSS and the correlation depends on halo mass or galaxy morphology. In this work, we use group and filament catalogues constructed from the SDSS DR12 to investigate the correlation between satellite systems and the large scale filaments they are located in. We find that the distribution of satellites is significantly correlated with filaments, namely the major axis of the satellite systems are preferentially aligned with the spine of the closest filament. Stronger alignment signals are found for the cases where the system away from the filament spine, while systems close to the filament spine show significantly weaker alignment.  Our results suggest that satellites are accreted along filaments, which agrees with previous works. The case of which away from the filament spine may help us to understand how the filament forms as well as the peculiar satellite distribution in the Local Universe.
\end{abstract}


\section{Introduction}\label{sec:intro}
According to the current cosmological paradigm, known as the Dark Energy Cold Dark Matter ($\Lambda$CDM) model, the early universe is nearly perfectly homogeneous but seeded with small density perturbations which, due to gravitational instability, grow increasing the density contrast of the universe \citep{Gunn1972ApJ...176....1G}.  Gravitational instability is responsible for halo collapse and, on large scales, the formation of the ``cosmic web'' \citep{Bond1996Natur.380..603B}, which consists of voids, sheets, filaments, and clusters.  Accordingly, matter collapses and forms dark matter haloes in a hierarchical fashion\citep{Davis1985ApJ...292..371D}: small haloes form first and these merge to form larger and larger structures. This process is accompanied by  mass flow on large scale towards regions of higher density. As halos merge, galaxies embedded inside halos merge too. The accreted and surviving smaller halos are referred to as ``subhalos'' and the galaxies that inhabit subhalos are called satellites. Therefore, understanding how the large-scale structure (hereafter, LSS) feeds subhalos toward halos is one of the keys to understanding how satellite systems form and how they affect the growth of larger haloes.    

Subhalos (satellite galaxies) are ideal tracers to study the mass distribution within halos. Recently, both numerical \citep{Agustsson2006ApJ...650..550A, Kang2005A&A...437..383K, Kang2007MNRAS.378.1531K, Libeskind2005MNRAS.363..146L, Wang2014ApJ...786....8W} and observational \citep{Brainerd2005ApJ...628L.101B, Yang2006MNRAS.369.1293Y, Wang2018ApJ...859..115W} studies have reported that subhalos (satellites) are preferentially distributed along their hosts' major axis. In addition, subhalos are often used as a bridge to study correlations between a halo (galaxy) and the cosmic web. Theoretical studies such as \cite{Libeskind2011MNRAS.411.1525L, KangWang2015ApJ...813....6K, WangKang2018MNRAS.473.1562W} have suggested that the spatial distribution of subhalos (satellite)  originates from their anisotropic accretion along a filamentary axis. Other studies, such as \citet{Tempel2015MNRAS.450.2727T}, have found that the satellites of host galaxies that are located in filaments, have an elongated distribution parallel to the filament axis. Similarly, \citet{Tempel2015A&A...576L...5T} show that galaxy pairs are also aligned with the cosmic web filaments.  Moreover, on the one hand, for the one halo term, observational and numerical studies confirm that the orientations of the major and spin axis of halos are correlated with their surrounding LSS \citep[such as][and reference within them]{Aragon2007ApJ...655L...5A, Hahn2007MNRAS.381...41H, Hahn2007MNRAS.375..489H, TempelLibeskind2013ApJ...775L..42T, WangKang2017MNRAS.468L.123W, WangKang2018MNRAS.473.1562W, 2018MNRAS.481.4753C, Welker2020MNRAS.491.2864W}. On the other hand,  for the two halo term, massive halos/clusters are correlated  on very large scales reaching up to $\sim100 \mpch$ \citep{2002A&A...395....1F, 2005ApJ...618....1H, 2012MNRAS.423..856S, 2017MNRAS.468.4502V, 2017ApJ...848...22X}. Interested readers can refer to two review papers by \cite{2015SSRv..193...67K} and \cite{2015SSRv..193....1J} for more details about the galaxy alignment.

In addition to the above, the study of satellite galaxy distribution as a whole (namely the distribution of the satellite system) has gained increased attention. On small scales, various confusing observations have been found in the Local Universe. \cite{LyndenBell1976MNRAS.174..695L} and \cite{Kunkel1976RGOB..182..241K} were the first to independently report, that satellite galaxies and distant globular clusters of the Milky Way align close to a common plane -- now termed the Vast Polar Structure (VPOS) \citep{Pawlowski2014ApJ...790...74P, Kroupa2005A&A...431..517K, Metz2008ApJ...680..287M, Metz2009ApJ...697..269M, Pawlowski2013MNRAS.435.2116P, Arakelyan2018MNRAS.481..918A}. Beyond the Milky Way, studies have found other cases and where  satellite galaxies appear to be preferentially aligned in significantly flattened planes: around M31 
\citep{Ibata2013Natur.493...62I, Conn2013ApJ...766..120C, Shaya2013MNRAS.436.2096S}, around Centaurus A \citep{Tully2015ApJ...802L..25T, Muller2016A&A...595A.119M, Muller2018Sci...359..534M}, around M101 \citep{Muller2017A&A...602A.119M}, and  around M83 \citep{Muller2018A&A...615A..96M}, although it must be noted that because of different biases, these may not all belong to an identical astronomical class of objects.

On larger scales, only a few studies have focused on the correlation between satellite systems and the LSS. By using the EAGLE simulation, \cite{Shao2016MNRAS.460.3772S} demonstrated that satellite planes tends to be aligned with the orientation of large scale structure. In observations, \cite{Libeskind2015MNRAS.452.1052L, Libeskind2019MNRAS.490.3786L} focused on the satellite systems in the Local Universe and reported that the normal directions of satellite systems are very closely aligned with the eigenvector of the velocity shear tensor, namely the direction of greatest compression of LSS. Related works such as \cite{2016MNRAS.463..222H, 2017MNRAS.468.4502V}, who used redMaPPer clusters to measure the alignment of the shapes of galaxy clusters, reported positive alignment signal.

In this paper, we investigate the correlation between satellite systems and LSS using the SDSS DR12 \citep{Eisenstein2011AJ....142...72E, Alam2015ApJS..219...12A}. Relative to the previous public release, DR12 adds one million new spectra of galaxies and quasars from the Baryon Oscillation Spectroscopic Survey (BOSS) over an additional 3000 deg$^2$ of sky. Owing to this advent of such large galaxy spectroscopic surveys, we are able to more accurately measure the 3D (rather than projected) distribution of satellites around centrals. This in turn will provide a statistically robust quantification of how satellites are distributed around their hosts.

The outline of the paper is follows. Section 2 presents the description of the data and method including the definition of satellite distribution, filament catalogue and alignment angle. In Section 3, we show the results of our analysis of the alignment angle between the orientation of the major axis of satellite distribution and their closest cosmic filament axis. Finally, we conclude and discuss our results in Section 4.

\begin{figure*}[!ht]
\plottwo{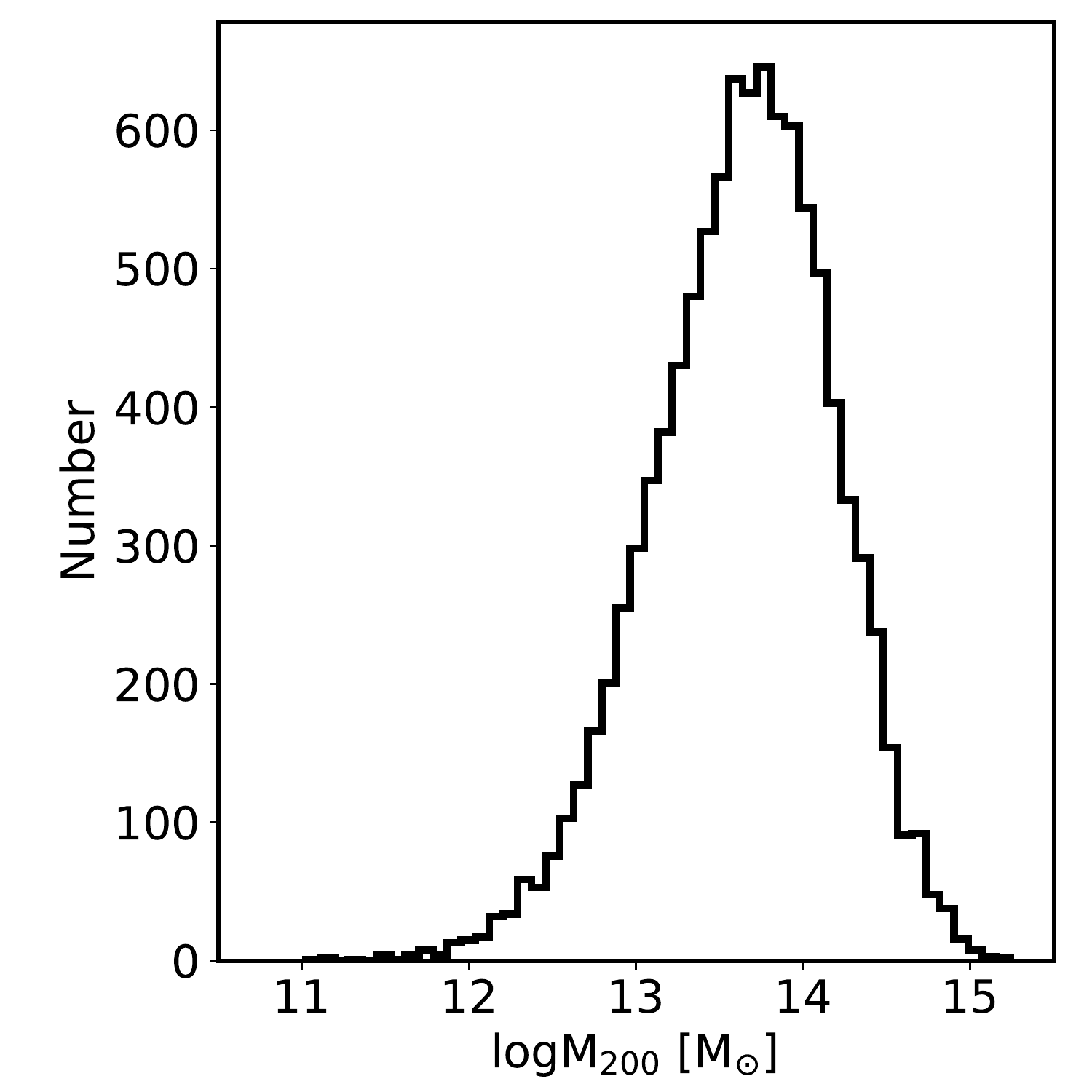}{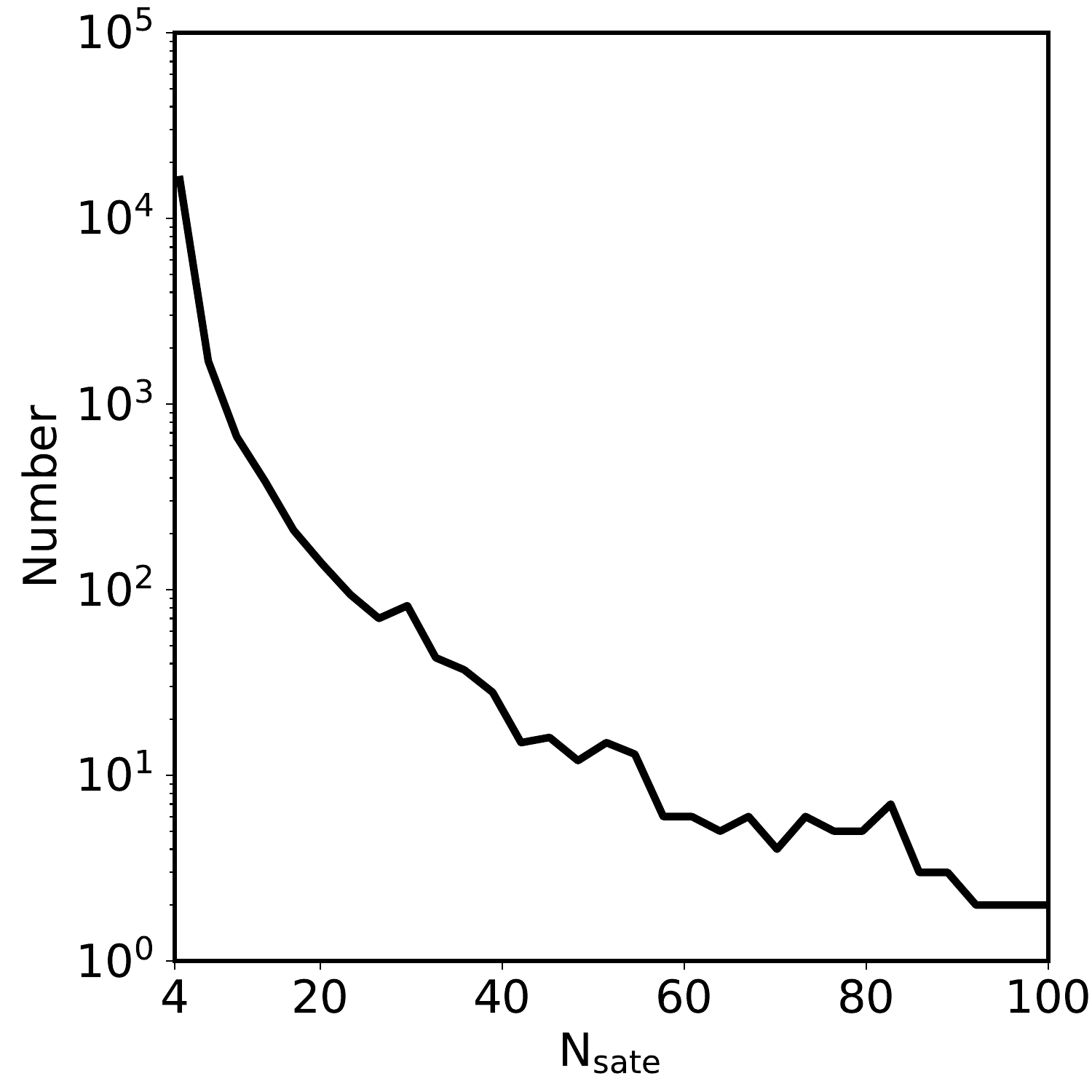}
\caption{Left panel: a histogram of group mass. Right panel: a histogram of $\nsate$, the number of satellites per group.}
\label{fig:sample}
\end{figure*}

\section{Data and Methodology}\label{sec:method}
\subsection{Observational Data}\label{sec:gg}
In this study, we employ the group catalog compiled by \cite{Tempel2017A&A...602A.100T}, in which galaxies from the spectroscopic sample of the SDSS data release 12 \citep{Eisenstein2011AJ....142...72E, Alam2015ApJS..219...12A} are grouped.  Groups are identified using a friends-of-friends (FoF) group finder, which was modified and designed specifically for flux-limited galaxy surveys, such as is the case here. The FoF group membership is refined in two main steps. First,  multi-modality analysis is used to split multiple components of groups into separate systems. The second step involves estimation of the virial radius and the escape velocity to exclude group members that are not physically bound to systems. The FoF group membership is refined to find subgroups. For more details, we refer the reader to \cite{Tempel2014MNRAS.438.3465T, Tempel2018A&A...618A..81T}.

The catalogue contains 88,662 groups with at least two members and 584,449 galaxies in total. Among them, we select groups with more than 5 members (i.e., 1 central plus 4 satellites). Our final fiducial catalogue consists of 10,087 groups with halo mass $\rm M_{200}$ roughly ranging from $10^{11}$ to $10^{15}$ $\msun$. In the left panel of Fig.~\ref{fig:sample}, we show the distribution of group mass. It is noted that throughout the paper we refer to the most luminous galaxy in each group as the central galaxy and all other members are regarded as satellites.

\subsection{Satellite Spatial Distribution}
In order to model the spatial distribution of satellites (hereafter referred to as the ``satellite ellipsoid'') in three dimensional space, we use the moment of inertia tensor whose definition is given by
\begin{equation}\label{equ:i_all}
I_{ij} = \sum_{k}^{N_{sate}}\frac{x_{i,k}x_{j,k}}{R^{2}_{k}},
\end{equation}
in which $i,j=\{0,1,2\}$ correspond to the Cartesian coordinate (for more detail about coordinate transformations, see \cite{Tempel2014MNRAS.438.3465T}, $x_{i,k}$ denotes the component $i$ of the position vector of satellite $k$ with respect to the central galaxy. $R_{k}$ is the distance between satellite $k$ and the central galaxy.  $\nsate$ is the number of satellites in the group and thus the number used to compute the satellite system's spatial distribution. In the right panel of the Fig.~\ref{fig:sample}, we show the number distribution of $\nsate$. Clearly, groups with a few satellites dominate the sample. The shape of the corresponding satellite ellipsoid is determined by the eigenvalues ($\lambda_{1}\geq\lambda_{2}\geq\lambda_{3}$) of $I_{ij}$. The length of each axis of the satellite ellipsoid are given by the square roots of the eigenvalues ($a=\sqrt{\lambda_1}$, $b=\sqrt{\lambda_2}$ and $c=\sqrt{\lambda_3}$). The thickness of the satellite ellipsoid is described using the minor-to-major axis ratio $c/a$ and the intermediate-to-major axis ratio $b/a$. The orientation of the satellite ellipsoid  is determined by the corresponding eigenvectors $e_{1}$, $e_{2}$ and $e_{3}$, respectively.  

\begin{figure}[!ht]
\plotone{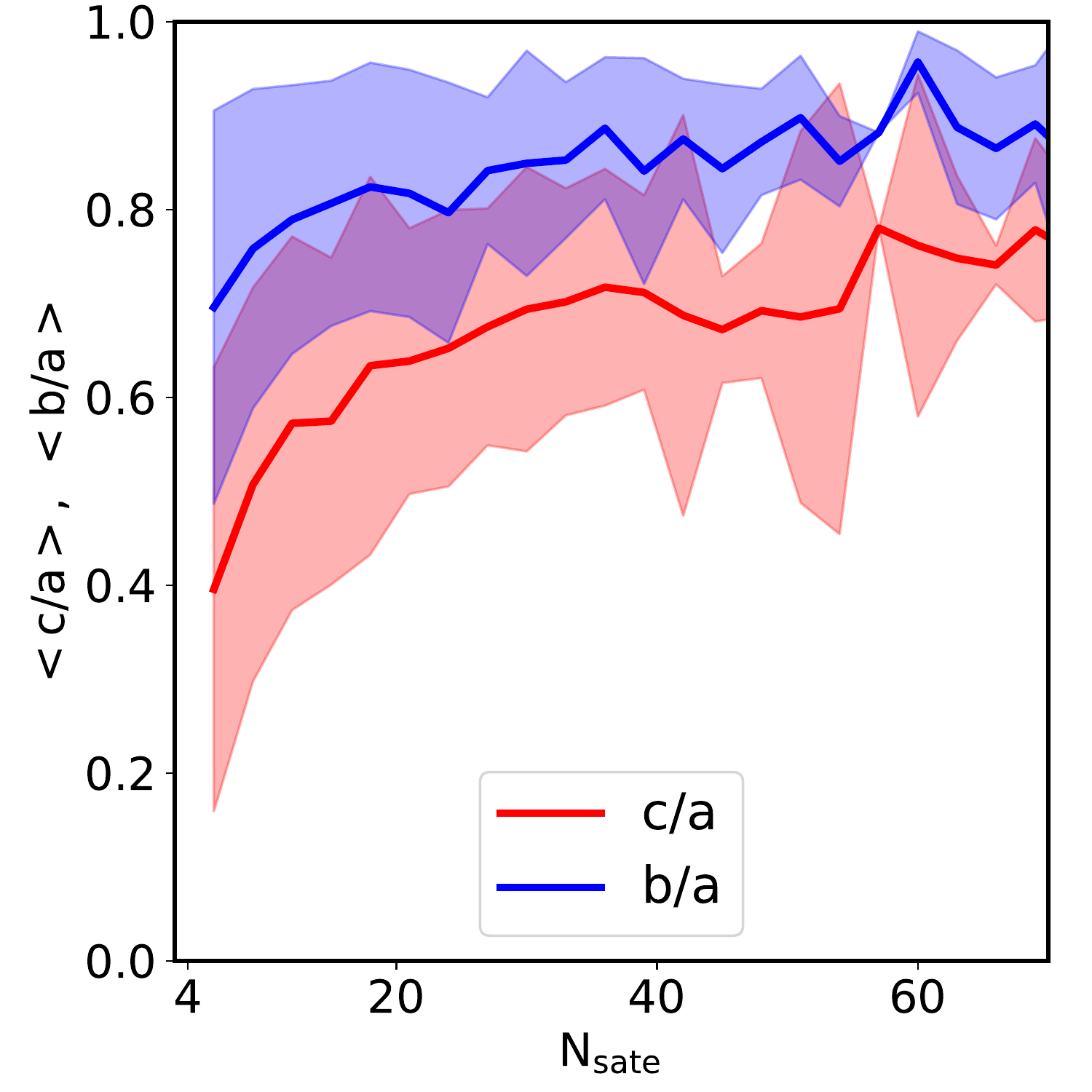}
\caption{The axis ratio, minor-to-major axis $\rm c/a$ in red and middle-to-major $\rm b/a$ in blue, as function of the number of satellites. Color bands show the $\rm 3\sigma$ standard deviation.}
\label{fig:sate_ca_ba}
\end{figure}

Geometrically, any ellipsoid in 3D space can be determined by at least four non-coplanar points (three satellites are always coplanar). Therefore, by including the central galaxy in the computation of $I_{ij}$, we may set the lower limit of $\nsate$ to 4. However, such shape determination is susceptible to stochastic Poisson noise. We will discuss how to avoid any Poisson noise in Section~\ref{sec:error}.  In Fig.~\ref{fig:sate_ca_ba}, we show how the $\nsate$ effects the determination of axis ratio $\rm c/a$ and $\rm b/a$. In general the smaller $\nsate$, the smaller the axis rations. The wide color bands show the $3\sigma$ spreads of axis ratios for a given $\nsate$: note a large spread for small values of $\nsate$ and relatively small spread for a large value of $\nsate$. We note that there is only one group in our sample with $\rm N_{sate}=57$ which leads to the zero standard deviation.

\subsection{Cosmic filaments}
The filaments are traced by applying an object point process with interactions (the Bisous process) to the distribution of galaxies in the spectroscopic galaxy sample \citep{Tempel2014MNRAS.438.3465T} from SDSS DR12. The filament finder is based on a robust and well-defined mathematical scheme that provides a quantitative classification consistent with the visual impression of the cosmic web. For more detailed descriptions of the algorithm, the readers can refer to \citep{Stoica2007JRSSC..56....1S, Tempel2014MNRAS.438.3465T, Tempel2016A&C....16...17T, Libeskind2018MNRAS.473.1195L}. For convenience, a brief summary is provided below.
Firstly, randomly oriented small cylindrical segments based on the positions of galaxies are used to construct a filamentary network by examining the connectivity and alignment of these segments. Then, a filamentary spine is extracted based on a detection probability and filament orientation. Finally, we repeat this process with a large number, a network of filaments with labelled coordinates, direction, and statistical significance emerges.

We note that, in this work, we only use central galaxies as tracers for filaments detection. Satellite galaxies are not used in the determination  of the filamentary network. This ensures that there is no intrinsic correlation between the satellite galaxy distribution and the filament orientation. The assumed radial scale for the extracted filaments is roughly $0.7 \  \rm  Mpc$. We classify a satellite system as ``close to filament'' if the distance of its central galaxy from the axis of the filament is less than $0.7 \  \rm  Mpc$. We term those galaxies who are more distant  than $0.7 \ \rm Mpc$  ``away from filament''. Note that not all filaments have a width of 0.7~Mpc. Indeed the filament finder returns a width for each detection. We examined the dependency of our results on the individual thickness of each filament and found that since most filaments have a width of around 0.7~Mpc \citep{Tempel2014A&A...566A...1T}, our results are robust to this choice.  We also use the distance normalized by the radius of the filament spine to define systems that are either in or not in the filament. We find these two methods return consistent results.

\subsection{Alignment Signal}\label{sec:angle}
In this work, we mainly focus on the orientation of the major axis of the satellite ellipsoid. Therefore throughout the paper the `orientation' of the satellite ellipsoid refers to the orientation (eigenvector) of the major axis of the satellite distribution. In order to examine (and quantify) if satellite ellipsoids are anisotropically distributed with respect to filaments, the probability density function (PDF, $\rm P(X)=N(X)/\langle N_{tot}(X) \rangle$) for both in 3D and (projected) 2D space are measured. 

For the 3D case we examine : 
\begin{equation}
  \cos(\theta)=|\rm e_{1} \cdot  \rm e_{fila}|,
\end{equation}
where $\theta$ is the angle between the orientation of the major axis of the satellite ellipsoid  $\rm e_{1}$  and the direction of the filamentary axis $\rm e_{fila}$, in which it is embedded.

Given the degeneracy of any inertia tensor eigenvector's positive direction, the alignment angle $\cos(\theta)$ is within $[0,1]$. For the case where the satellite ellipsoid  is randomly oriented relative to the filament axis, the expectation of $\langle| \cos(\theta)|\rangle$ is $0.5$.

For the 2D on-sky case, we measure the alignment signal by projecting the major axis of the satellite ellipsoid and the filament axis onto the plane of the sky. The alignment angle $\theta_{\rm SF}$ is the difference between position angle of the satellite ellipsoid $\theta_{\rm S}$ and the position angle of the filament $\theta_{\rm F}$, namely $\theta_{\rm SF}=|\theta_{\rm S}-\theta_{\rm F}|$. Following the method used in \cite{Zhang2013ApJ...779..160Z}, we compute the position angle $\theta_{\rm SF}$ with respect to the closest filament at the location of the central galaxy $\boldsymbol{x}$, written as:

\begin{equation}\label{position_angle}
\theta_{\rm SF} = \arctan\left[ \frac{\Delta \alpha \, \cos \delta_{\bx}}
{\Delta \delta}\right],
\end{equation}
\\
where $\Delta \alpha$ and $\Delta \delta$ are the right ascension and declination differences, respectively. $\delta_{\bx}$ is the declination of central galaxy. The position angles are measured at the location of central galaxy $\bx$ and at the location of the filament which can be offset. Note that the alignment angle $\theta_{\rm SF}$ is restricted to be within $[0^{\circ},90^{\circ}]$. For the case that satellite ellipsoid  are randomly distributed with respect to the filament's axis, the expectation is $\theta_{\rm SF}=45^{\circ}$.

The possible alignment signal is calculated using the kernel density estimation. \cite{Tempel2014MNRAS.437L..11T} suggested that the kernel density estimation is better than a simple histogram and is more representative of the underlying probability distribution. For $\rm |\cos(\theta)|$=1 or $\rm \theta=0^{\circ}$ implies that the satellite ellipsoid  are ideally distributed along the filament' axis (hereafter referred to as ``alignment''), while $\rm |\cos(\theta)|$=0 or or $\rm \theta=90^{\circ}$ implies that satellite ellipsoid  ideally distributed perpendicular to the orientation of the filament (hereafter termed ``anti-alignment'', not to be confused with parallel but pointing in opposite directions).

\begin{figure*}[!ht]
\plottwo{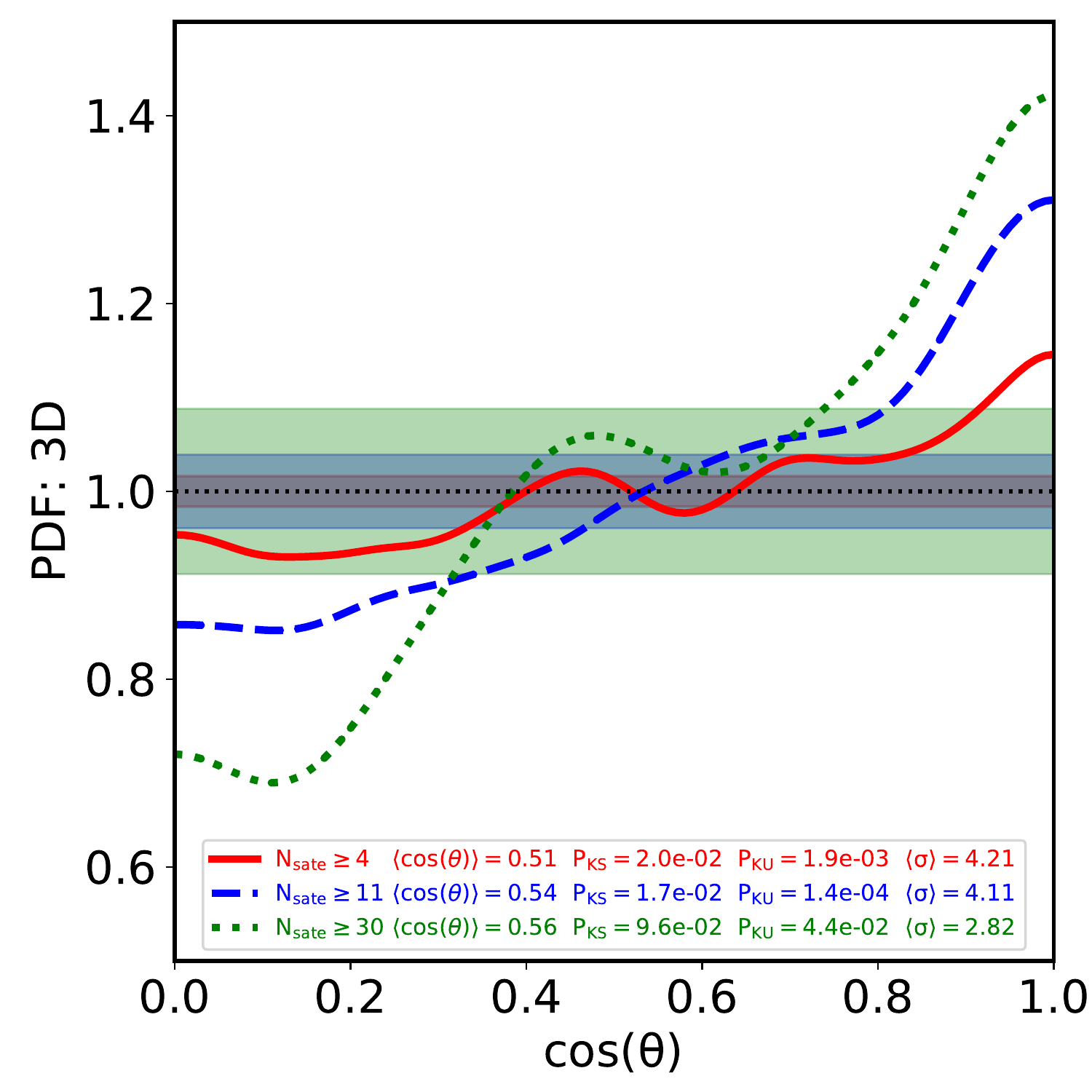}{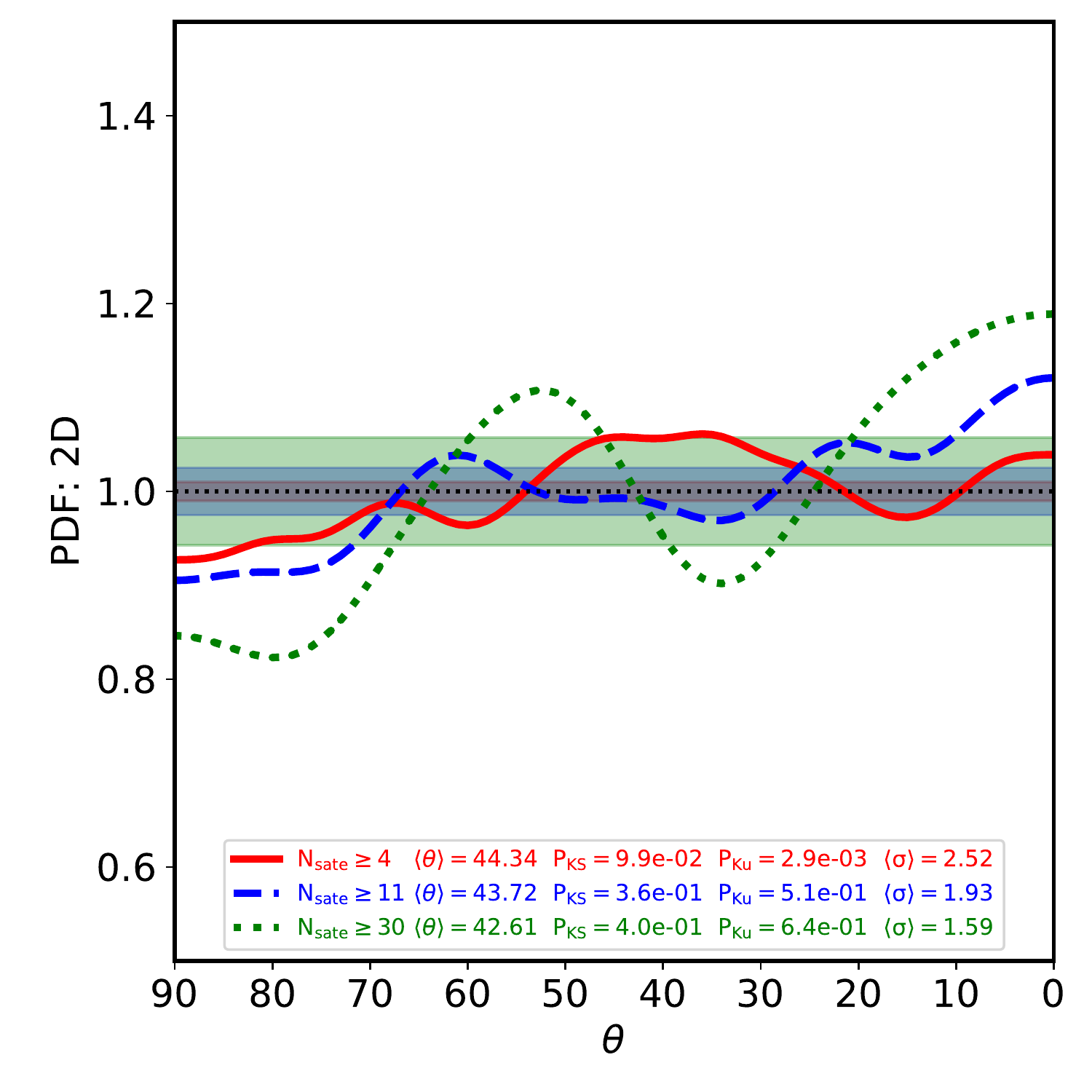}
\caption{The probability distribution function of the alignment between filaments and the major axis of satellite distributions. On the left panel we show $\cos(\theta)$ and on the right panel  the projected angle $\theta$. Different lines separate groups of different richness, namely with different $\nsate$. The horizontal black dotted lines represent a random, uniform distribution. The color bands show the $3\sigma$ spread about  the mean obtained by randomizing satellite position in 3D (left panel) and in projection (right panel) 10,000 times. The mean alignment angle, $\rm \pks$ of the Kolmogorov-Smirnov (KS) test, the associated probability $\rm \pku$ of the Kuiper test, and the average significance of the alignment signal are shown in the legend box with corresponding colors.}
\label{fig:f1}
\end{figure*}

\begin{figure}[!ht]
\plotone{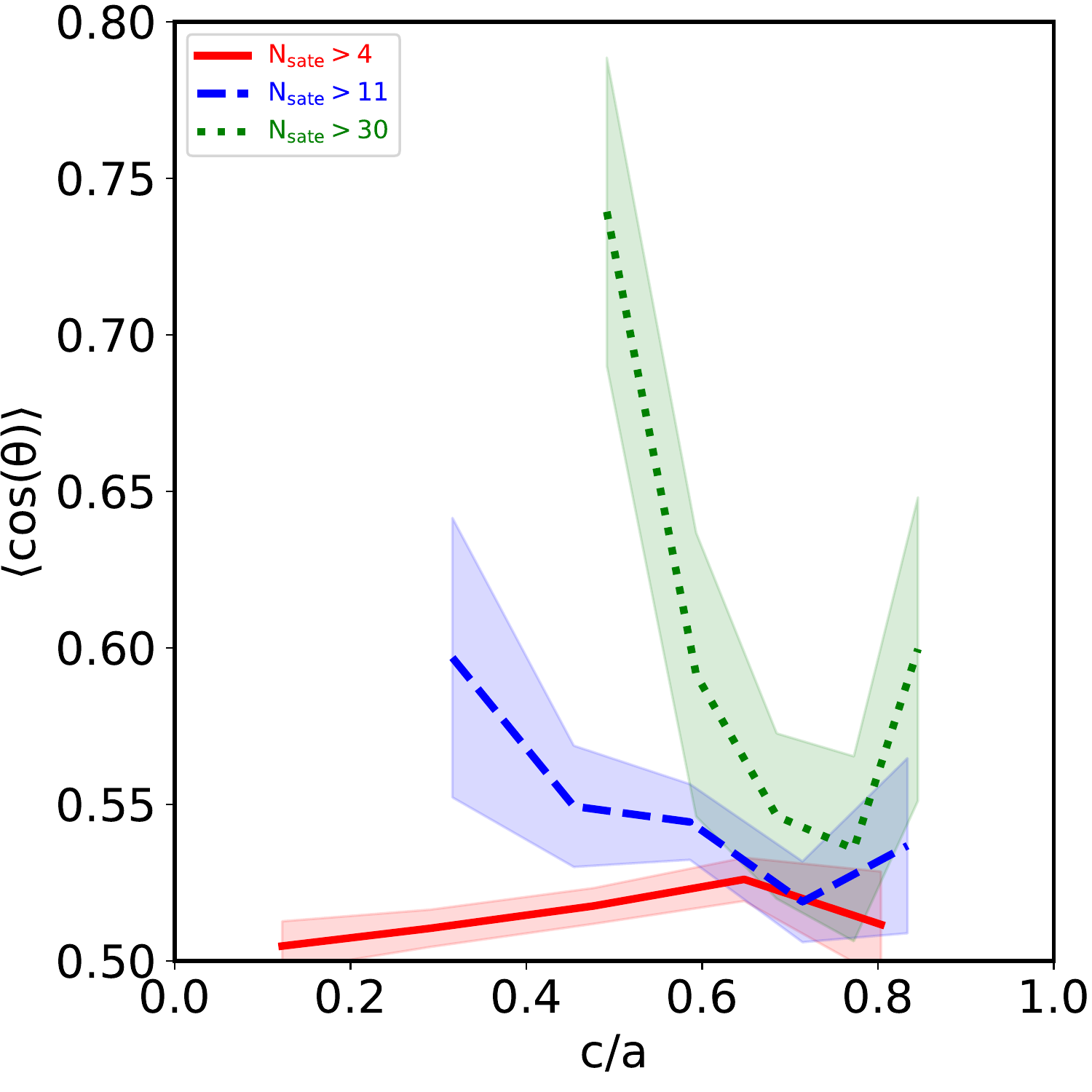}
\caption{The dependence of the alignment signal $\langle\cos(\theta)\rangle$ on the axial ratio
$c/a$ is examined. Three sub-samples same as Fig.~\ref{fig:f1}, namely systems with $\nsate\geq4$ (red solid line), $\nsate\geq11$ (blue dashed line) and $\nsate\geq30$ (green dotted line).}
\label{fig:add_ca}
\end{figure}

\begin{figure*}[!ht]
\plottwo{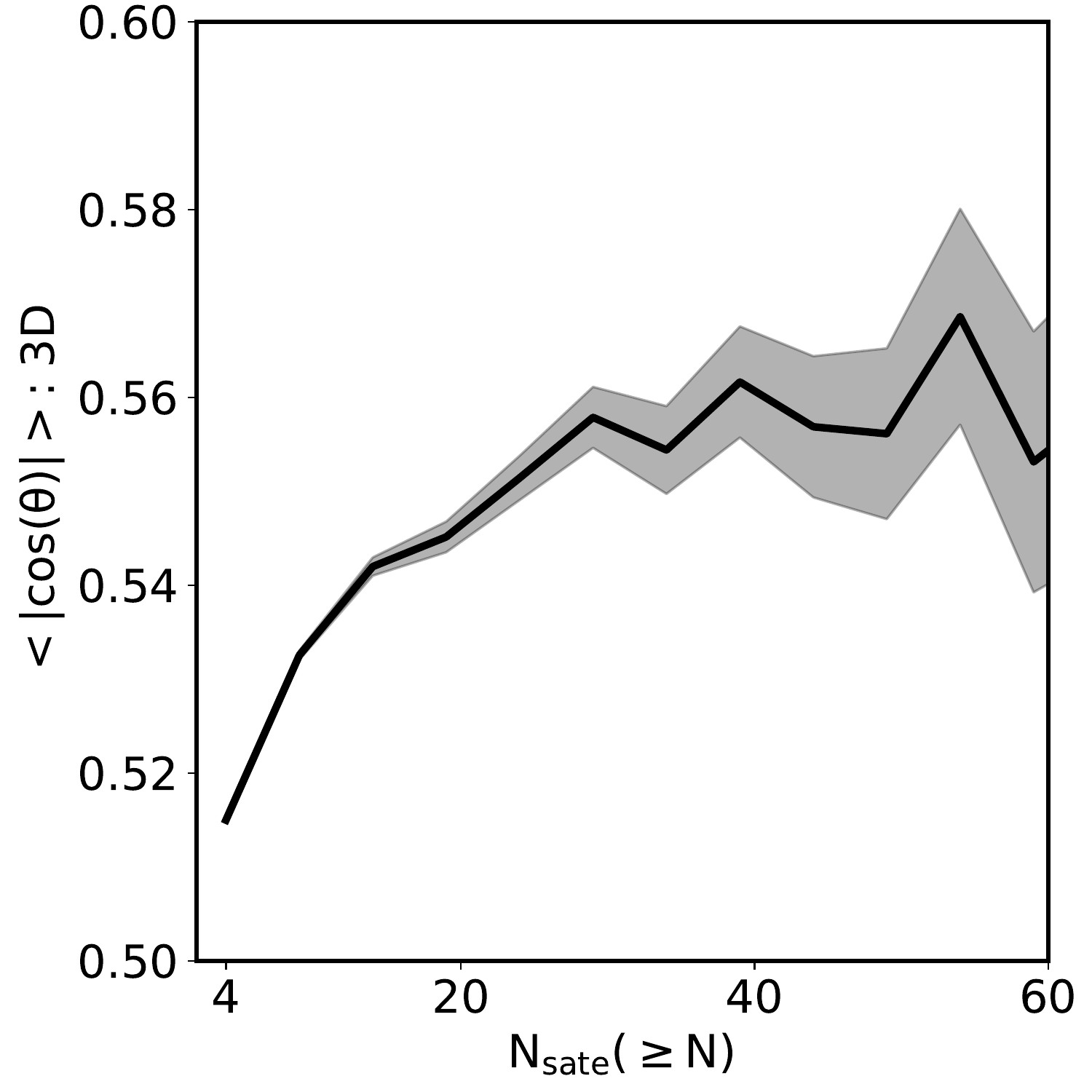}{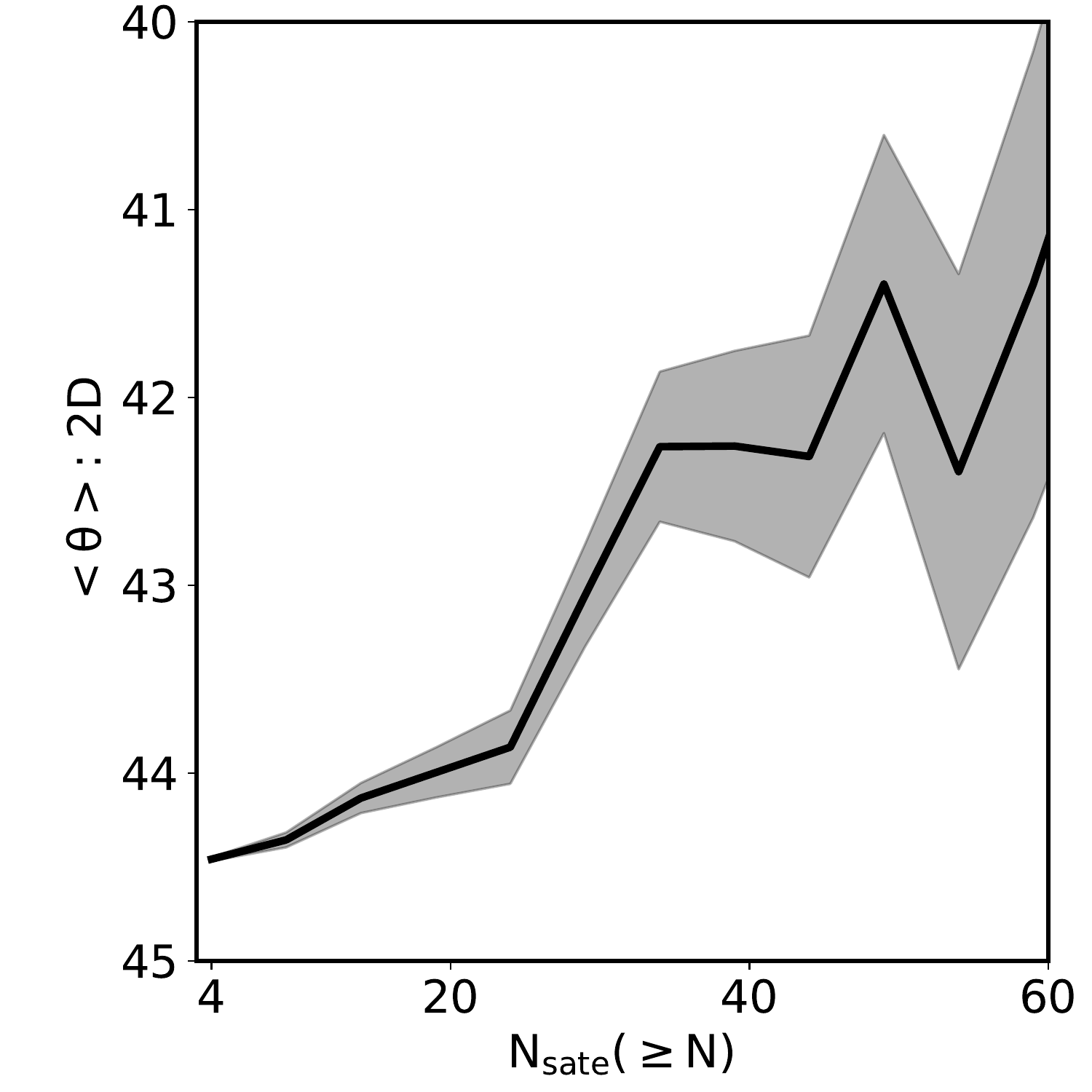}
\caption{The alignment angle as a function of the number of the satellites $\nsate$ both in 3D (left panel) and 2D (right panel). A clear increasing tendency can be seen for low $\nsate$ and become roughly independent with $\nsate$ at $\nsate\geq30$. The color brand indicate the $1\sigma$ uncertain obtain from 10,000 random sample.}\
\label{fig:nsate_angle}
\end{figure*}

\begin{figure}[!ht]
\plotone{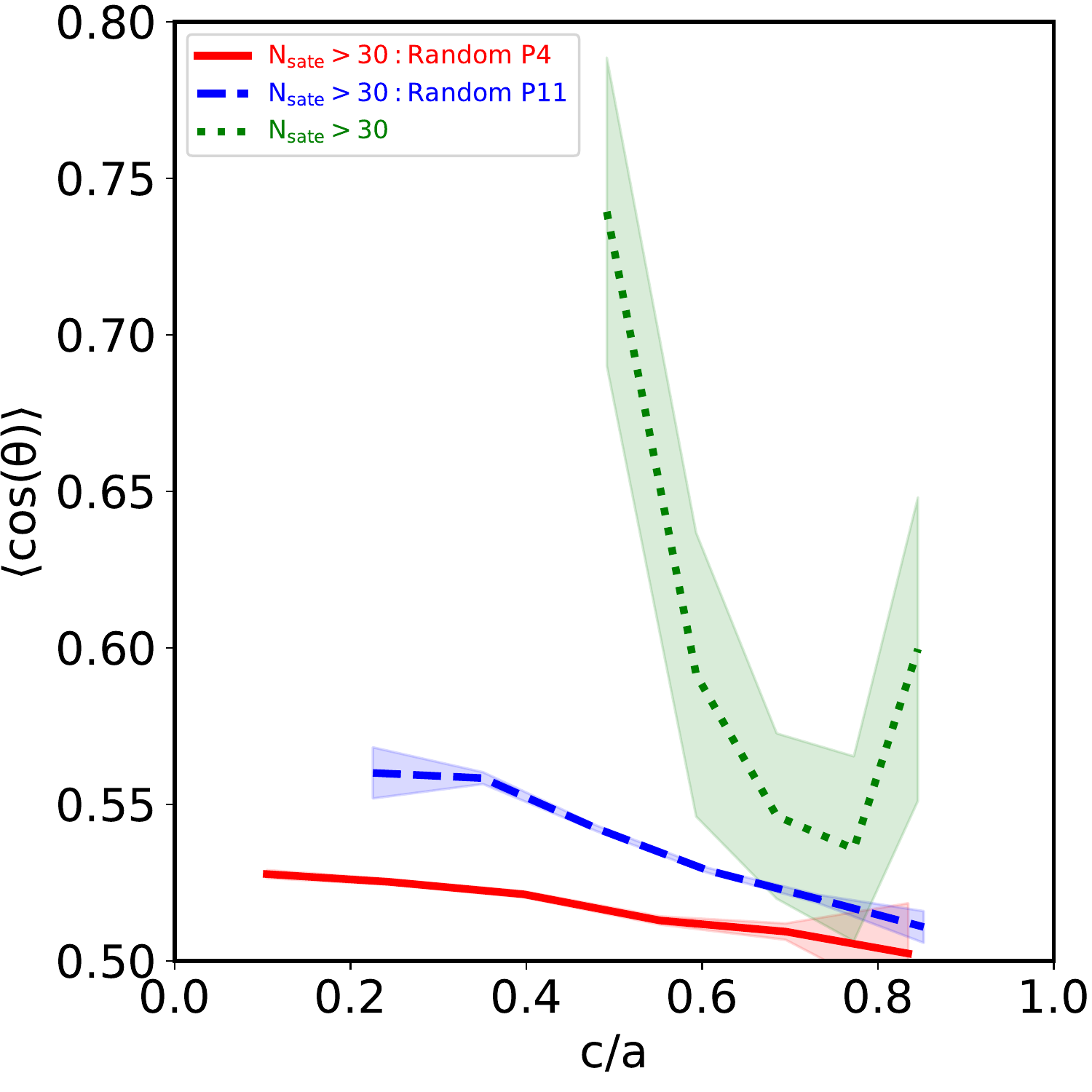}
\caption{Same as Fig.~\ref{fig:add_ca}, but show two different sub-samples which are constructed from groups with $\nsate\geq30$. For these two different sub-samples, we randomly picked either 4 (red solid line) or 11 (blue dashed line) satellites from those groups which contain more than 30 satellites. This process was performed 10,000 times.}
\label{fig:add_ca2}
\end{figure}

\subsection{Error Estimation}\label{sec:error}
The orientation of filaments and the estimation of satellite ellipsoids are affected by redshift-space distortions. Our analysis thus includes a measure of the alignment signal in the plane of the sky (see 2D case above in Section~\ref{sec:angle}).  

Furthermore, the number of satellites $\nsate$ affects the determination of axes ratios found when fitting satellite ellipsoids. The statistical significance of any  or axis ratio measured for a group of size $\nsate$ can be gauged via a simple Monte-Carlo test. For each group, the radial distribution of group members (satellites) is kept constant but their angular distribution is randomized. This test is performed 10,000 times per group. Each time, the inertia tensor and it's eigenvectors and eigenvalues are determined for the randomized satellite distribution. As such, a mean alignment between the randomized satellite distribution and the filament can be estimated. The statistical significance of any measured alignment signal can be estimated in terms of the standard deviation of such randomized trials.

\begin{figure*}[!ht]
\plottwo{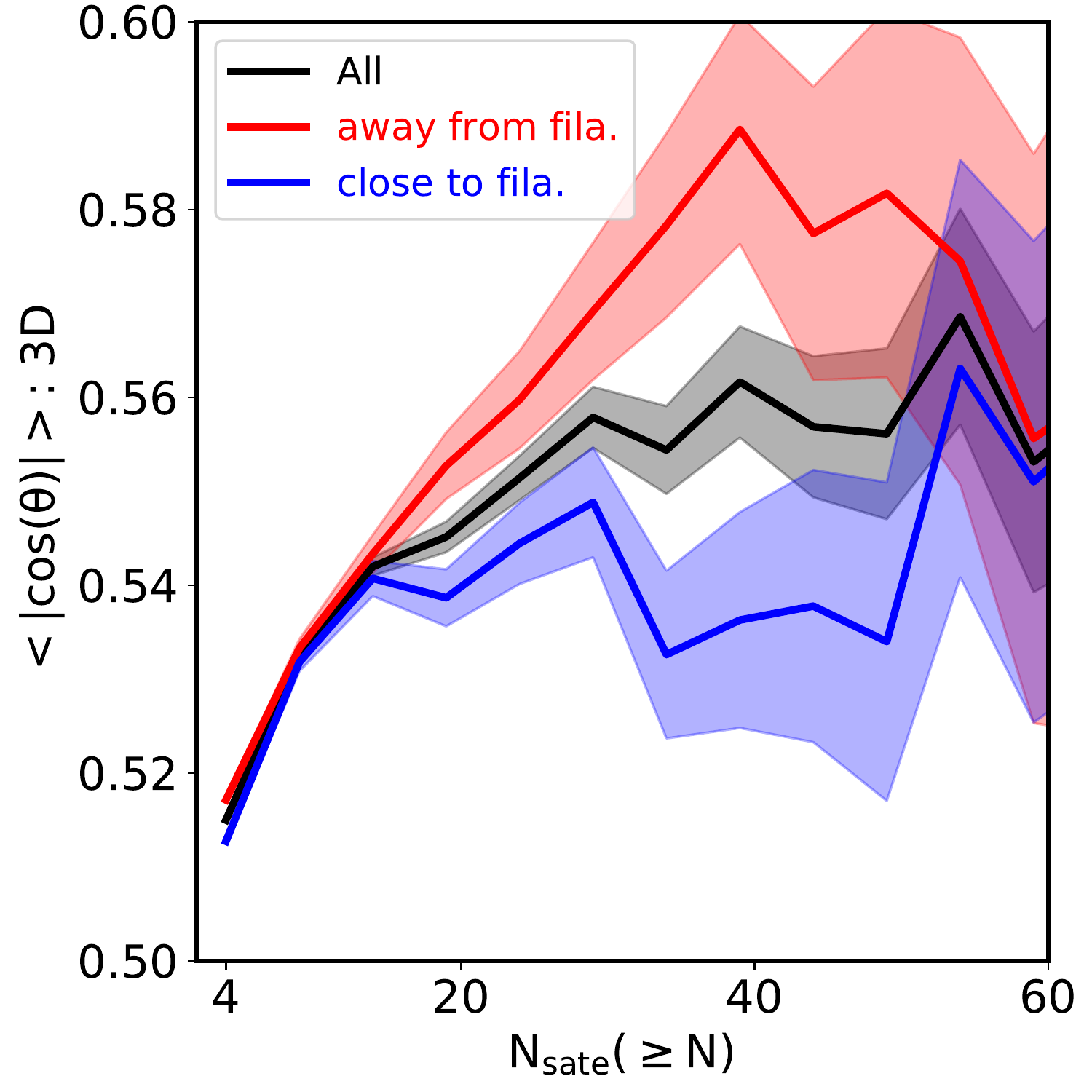}{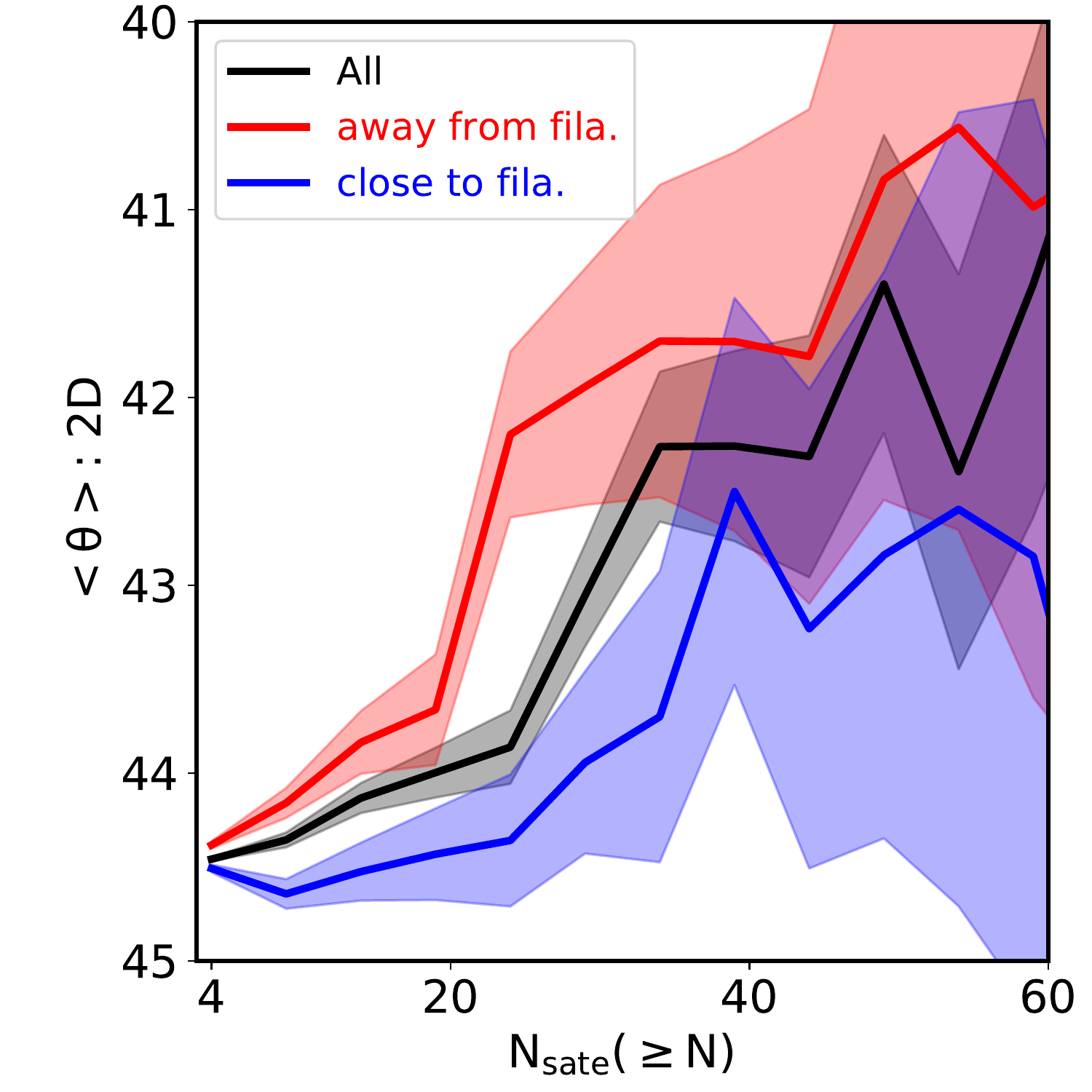}
\caption{Same as Fig.~\ref{fig:nsate_angle}, but the alignment signals are examined for group either ``close to'' (in blue) or ``away from'' (in red) the filaments. The color brand indicate the $1\sigma$ uncertain obtain from 10,000 random sample, respectively.}
\label{fig:dist}
\end{figure*}

\section{Results}\label{sec:results}

The alignment between the orientation of the major axis of the satellite ellipsoid and their respective filament axis, expressed as a probability distribution of the angle (in 2D) or cosine of the angle (in 3D) is presented in Fig.~\ref{fig:f1}. The left panel shows the alignment signal examined in 3D and the measured alignment signal in the plane of the sky (2D) is shown in the right panel. The group sample is divided according to the number of satellites that constitute it. The black dotted line indicates the expectation if satellite ellipsoids are randomly distributed with respect to the filament axis (namely a uniform distribution). The corresponding color bands show the $3\sigma$ fluctuation expected given a sample size  (see Section.~\ref{sec:error} for more details). In the bottom of each panel, we show the mean values of the alignment angles, either $\cos(\theta)$ in the left panel or $\theta$ in the right panel, the statistical significance $\langle\sigma\rangle$ (namely how far the mean angle is from $\cos\theta=0.5$ for 3D or 45$^{\rm o}$ for 2D), the $p$-value $\pks$ of the Kolmogorov–Smirnov (KS) test , and the associated probability $\rm \pku$ of  the Kuiper test. The the statistical significance $\langle\sigma\rangle$ and the KS (Kuiper) test are performed to quantify the chance that such distributions could arise from a uniform distribution. Note that a high value of average significance and a low value of $\pks (\pku)$ indicates a statistically robust alignment signal.

In Fig.~\ref{fig:f1}, the red solid lines denote the full sample that includes all groups (with $\nsate \geq 4$), while the colored dotted and dashed lines denote the alignment for groups chosen according to satellite number. When considering all groups irrespective of satellite number, Fig.~\ref{fig:f1} shows that their major axis is well aligned with the filament axis in the 3D case (left panel) with a mean value of $\meancosangle=0.51$, a high value of statistical significance ($\sim 4\sigma$), and low value of $\pks=2.0\times10^{-2}$ ($\pku=1.9\times10^{-3}$). However, for the 2D case shown in the right panel, the alignment  becomes more uniform with a mean value of $\meanangle=44.34^{\circ}$, a statistical significance of $\sim 2.52\sigma$. The large sample size in this case results in a statistically more robust signal, albeit the lower significance in the 2D case is clearly due to projection effects washing out the signal. 

Two additional examinations are performed in Fig.~\ref{fig:f1}. The groups are divided in to two sub-samples according to $\nsate$ (blue dashed lines for groups with $\geq$11 and green dotted lines for those of with $\geq30$ satellites). The (seemingly arbitrary) choice of 11 and 30 is motivated as follows: 11 is the number of Milky Way classic satellite galaxies (whose normal is not far from the local filamentary direction \citep{Libeskind2015MNRAS.452.1052L}. Thus this sample allows us to compare, to some extent, the Milky Way's situation with groups that have the same number of satellite. The cut at 30 satellites is also examined because according to Fig.~\ref{fig:sate_ca_ba}, this is the value where the $c/a \ (b/a)$ become roughly independent of $\nsate$. In both these cases we see an increase in the alignment signal but, owing to the smaller sample size, an accompanying reduction in the statistical significance. As is expected, all signals are further reduced in the 2D alignment signal. Overall, we conclude that \textit{the major axes of satellite ellipsoids are preferentially aligned with the axes of their filaments.}

Despite the weaker statistical significance, we wish to  assess if the stronger alignment signal seen for richer groups is real or due to a systems with better defined axis ratio or satellite number. In order to do so, we  examine the alignment signal as a function of satellites ellipsoid axis ratio $c/a$, shown in Fig.~\ref{fig:add_ca}. The same three sub-samples as in Fig.~\ref{fig:f1} are checked. It is clear that groups with $\nsate\geq 30$ and $\nsate\geq11$ show an obvious trend: as the axis ratio approaches sphericity, the alignment signal weakens. Furthermore, in Fig.~\ref{fig:nsate_angle}, we examine how $\nsate$ affects the alignment signal,  by examining the mean angle as function of $\nsate$ for both in the 3D and 2D case. We find that for low value of $\nsate$, the $\meancosangle$ increases with $\nsate$, while for $\nsate$ higher than $\sim30$, the $\meancosangle$ become independent with $\nsate$. We note that there exists the possibility that this trend is due to limited sampling, namely that when only a few satellites are chosen that this may bias the $\meancosangle$-$\nsate$ trend. To test this effect, we randomly select either 4 or 11 satellites from groups with more than 30 satellites, compute their axis ratio and alignment signal. The results are shown in Fig.~\ref{fig:add_ca2}. The alignment signal of two random samples are comparable with groups that contain the same number of satellites (the corresponding color line in Fig.~\ref{fig:add_ca}), indicating that the number of satellites indeed does have an effect on the alignments. If the satellite systems are in fact richer than what we observe, they are more likely to be well aligned.

In Fig.~\ref{fig:dist}, we examined whether the distance between group and filament affects the alignment signal. We divide the groups according to the distance from the central galaxy to the filament spine: either they are close to the spine of the filament (within a distance of $d_{\rm group} < 0.7$ Mpc) or they are further. The alignment signal for the $d_{\rm group} \geq 0.7$ Mpc case is shown in red solid lines, while for the $d_{\rm group} < 0.7$ Mpc sample is shown in the blue solid lines.  We find that the satellite alignment in groups with $d_{\rm group} \geq 0.7$ Mpc is stronger than those whose central galaxies are close to the filament spine, in agreement with \cite{2018A&A...613A...4W}.  In both 3D and 2D cases however there is a statistically significant signal.

\section{Conclusion and Discussion}\label{sec:con_dis}
In this work, using groups and filaments constructed from SDSS DR12, we investigate  the  correlation  between  satellite systems and the LSS, namely the alignment between the major axis of satellite distributions and their closest filament axis. We have conducted our analysis using 3D positions as well as projected 2D positions on the sky. Our main results can be summarized as follow:

\begin{itemize}
    \item the major axes of satellite ellipsoids are preferentially aligned with the axes of the filaments they inhabit with high significance. Such an alignment is statistically significant regardless of the number of satellites that compose each group.
    \item Galaxies that reside close to the spine if the filament they inhabit show a slightly weaker alignment compared to galaxies that are a bit further away whose alignment signal is stronger.
\end{itemize}

The main result, shown in the Fig.~\ref{fig:f1}, are consistent with the published theoretical results of \cite{Shao2016MNRAS.460.3772S} which are based on $\Lambda$CDM simulations of galaxies with Milky Way halo mass range. They also agree with the observational work of \cite{Libeskind2015MNRAS.452.1052L, Libeskind2019MNRAS.490.3786L}  based on $\sim$ 10 systems in the Local Universe. Our results, for the first time, extend this correlation  between  satellite systems and the LSS to a large sample of observations.

Partly, our results concern the relationship between the number of satellites and the distribution of angle that the long axis of a fitted ellipsoid makes with the filament such a group resides in namely $\meancosangle$-$\nsate$, shown in the Fig.~\ref{fig:nsate_angle}.  There are two possible explanations. Theoretically, on the one hand, both numerical simulations \citep[e.g.,][]{vanHaarlem1993ApJ...418..544V, Aragon2007ApJ...655L...5A, Hahn2007MNRAS.381...41H, Hahn2007MNRAS.375..489H, Zhang2009ApJ...706..747Z, KangWang2015ApJ...813....6K}(for summary tables, see \cite{Forero-Romero2014MNRAS.443.1090F} and \cite{Wang2018ApJ...866..138W}) and observations
\citep[e.g.,][]{Zhang2013ApJ...779..160Z, Faltenbacher2009RAA.....9...41F, Godlowshi2010ApJ...723..985G, Godlowshi2011IJMPD..20.1643G, Godlowshi2012ApJ...747....7G, Paz2011MNRAS.414.2029P, Skielboe2012ApJ...758L..16S, Li2013ApJ...768...20L}
show that the shape of a dark matter halo (used as a proxy for the shape of a satellite distribution) to be aligned with their surrounding large scale structure, and the strength of alignment depends on mass. Furthermore, the more massive the halo (or group), the more ellipsoidal the halo shape \citep{Allgood2006MNRAS.367.1781A} and hence stronger the alignment \citep{Benson2005MNRAS.358..551B, Yang2006MNRAS.369.1293Y, Wang2018ApJ...859..115W}. On the other hand, according to the $\Lambda$CDM power spectrum, the more massive the halo, the more satellite galaxies or subhaloes it will contain. Thus, if the distribution of satellite galaxies is able to trace the halo shape, the larger $\nsate$, the stronger the alignment signal. 

However, based on the samples used in this work, we found the $\meancosangle$-$\nsate$ is influenced by sampling effects. The systems with small $\nsate$ have less well-constrained orientations since the shape is determined with fewer satellites. In a related study, \cite{2016MNRAS.463..222H} discussed the dependence of satellite-central alignment on the cluster richness. We leave theoretical predictions regarding the $\meancosangle$-$\nsate$ to future simulation studies.

Whats more, the accretion of satellite galaxies has been found to be a strongly anisotrpic process. Some of these studies \citep[e.g.,][]{Benson2005MNRAS.358..551B, Libeskind2005MNRAS.363..146L}  have shown that satellite galaxies are found to be accreted anisotropically during the formation history of halos (groups), and the current distribution of satellite galaxies retains the memories of the primordial accretion. \cite{KangWang2015ApJ...813....6K} suggested that satellite galaxies tend to be accreted along the direction of the major axis of the host halo, while \cite{Libeskind2015MNRAS.452.1052L, KangWang2015ApJ...813....6K, Kubik2017MNRAS.472.4099K, Libeskind2011MNRAS.411.1525L} found that satellite galaxies are accreted preferentially along the filament axes. The current satellite-central alignment is the result of the competition between primordial accretion and non-linear evolution inside the halo \citep{KangWang2015ApJ...813....6K, Wang2018ApJ...866..138W}. These theoretical works predict (or at least suggest to high degree) that the orientation of satellite ellipsoids should be correlated with the LSS, shown here observationally.

The anisotropic infall of satellite galaxies onto host halos is often invoked as a possible explanation of the formation of the highly flattened distribution satellite galaxies such as in the case of the MW as well as other galaxies in the Local Universe such as M31 and Centaurus A \citep[e.g.,][]{Libeskind2005MNRAS.363..146L}.  However, it is unclear if the numerical studies predict enough anisotropy to explain such a phenomenon \citep[i.e. see][]{Libeskind2015MNRAS.452.1052L, KangWang2015ApJ...813....6K}. Therefore, it can explain both the thin and corotating nature of such systems \citep{Pawlowski2013MNRAS.435.2116P}. Our results indicate that there is a stronger alignment  for groups located further from the filament spine. \cite{2018A&A...613A...4W} also found the opposite trend for the spatial distribution of satellite galaxies relative to their closest filaments: their alignment is dominated by centrals at large distance and fades away in its vicinity. We suspect this is because there is shell crossing and non-linear behavior in the formation of the systems that are deeper in the filamentary potential. It remains to be seen if numerical simulations can falsify this hypothesis. Either way, the fact that the alignment signal depends on filament proximity may shed light on the orientation of such peculiar satellite distribution formed in the Local Universe.

\acknowledgments
{\bf Acknowledgement:}
PW, NIL, XK and QG acknowledge support from the joint Sino-German DFG research Project ``The Cosmic Web and its impact on galaxy formation and alignment'' (DFG-LI 2015/5-1, NSFC No. 1186131006). NIL acknowledges financial support of the Project IDEXLYON at the University of Lyon under the Investments for the Future Program (ANR-16-IDEX-0005). NIL, MSP, and PW also thank the DAAD for PPP grant (57512596). ET was supported by ETAg grant IUT40-2 and by EU through the ERDF CoE TK133 and MOBTP86. XK acknowledges fininaical supported by the NSFC (No. 11825303, 11333008), the 973 program (No. 2015CB857003, No. 2013CB834900). QG acknowledges fininaical support of  Shanghai Pujiang Program (No.19PJ1410700).

 \end{CJK*}
\end{document}